# Can Multiple-Choice Questions Simulate Free-Response Questions?

Shih-Yin Lin and Chandralekha Singh

*Department of Physics and Astronomy, University of Pittsburgh, Pittsburgh, PA, 15260 USA*

**Abstract.** We discuss a study to evaluate the extent to which free-response questions could be approximated by multiple-choice equivalents. Two carefully designed research-based multiple-choice questions were transformed into a free-response format and administered on final exam in a calculus-based introductory physics course. The original multiple-choice questions were administered in another similar introductory physics course on final exam. Findings suggest that carefully designed multiple-choice questions can reflect the relative performance on the free-response questions while maintaining the benefits of ease of grading and quantitative analysis, especially if the different choices in the multiple-choice questions are weighted to reflect the different levels of understanding that students display.

## INTRODUCTION

When it comes to assessing students' learning in physics, there is always concern about the format of the assessment tool. While a multiple-choice (MC) test provides an efficient tool for assessment because it is easy to grade, instructors are often concerned when using it because a test in a free-response format facilitates a more accurate understanding of students' thought processes. In addition, free-response questions allow students to get partial credit for displaying different extent of understanding of the subject matter tested, which is appreciated by many instructors and students. Thus, there appears to be a trade-off between the two assessment tools. If the instructors choose to implement a multiple choice test, they often feel that they are completely sacrificing the benefits that the free-response questions could provide.

Research indicates that the difficulties students have related to a given topic can be classified into relatively few categories. If the choices in the MC questions are designed carefully to reflect the common difficulties students have, it is possible that the multiple choice questions will faithfully reflect the performance on the free-response questions while maintaining their benefits of ease of grading and comparison of classes taught using different instructional approaches. Here, we present a study designed to investigate the relation between students' performance on quantitative free-response and MC questions. We converted two research-based MC questions into free-response format and administered them on final exam in a calculus-based introductory physics course (course A). The original MC questions were administered in an equivalent introductory physics course (course B) on final exam. Students' performance in two different courses is compared. Moreover, we investigate the correlation between students' actual performance on the free-response questions and a "simulated" multiple-choice performance had the problems been given in the MC format in course A.

## METHODOLOGY

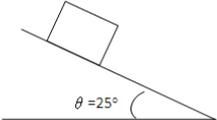

FIGURE 1. Problem Statement for Question 1.

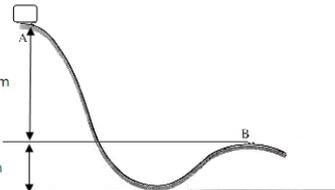

FIGURE 2. Problem Statement for Question 2.

Figures 1 and 2 present the MC questions that were administered in this study. Question 1 concerns an object at rest on an inclined plane. Students were asked to find the magnitude of the static friction acting on the object, which is equal to $mg \sin\theta$ according to Newton's 2nd Law. Research [1,2] suggests that many students struggle with this question because they believe that the magnitude of static friction ($f_s$) is always equal to its maximum value, the coefficient of static friction ($\mu_s$) times the normal force ($F_N$). This notion is not valid for this question because the maximum value of static friction exceeds the actual frictional force needed to hold the object at rest. Other student difficulties found from a pilot study include the confusion between static and kinetic friction, the challenge in decomposing the force correctly, etc. The alternative choices in the MC question are designed to reflect these difficulties. Table 1 presents the choices in algebraic form where all symbols have their standard meaning (actual choices were numerical).

TABLE 1. The algebraic form for the choices in question 1 and the different scores assigned in the "weighted multiple-choice" simulation. The correct answer is indicated by the shaded background.

| Choice | Question 1 | Score |
|---|---|---|
| (a) | $mg\sin\theta$ | 1.0 |
| (b) | $\mu_k mg\cos\theta$ | 0.3 |
| (c) | $\mu_s mg\cos\theta$ | 0.5 |
| (d) | $mg\cos\theta$ | 0.2 |
| (e) | none of the above | 0.0 |

Question 2 concerns a roller coaster cart on a frictionless track. The question asks for the normal force acting on the cart when it goes over a hump, which can be solved by using the principles of the conservation of mechanical energy and Newton's 2nd law in the non-equilibrium situation (with centripetal acceleration involved). Previous research [3] indicates that a common difficulty introductory physics students have is that they think of a non-equilibrium situation which involves the centripetal acceleration as an equilibrium situation by treating the centripetal force as an additional force. The correct use of the centripetal acceleration and Newton's 2nd Law in this question should yield $N - mg = -\frac{mv^2}{r}$. However, students who treat it as an equilibrium question and believe that the centripetal force is an additional force obtain an answer of the type $N - mg - \frac{mv^2}{r} = 0 \Rightarrow N = mg + \frac{mv^2}{r}$, which has an incorrect sign. In addition to this common mistake, a pilot study indicates that some students incorrectly believe that the normal force is equal to the gravitational force ($N = mg$) without contemplating the centripetal acceleration. On the other hand, there are also students who completely skip the gravitational force and claim that $N = \frac{mv^2}{r}$. Moreover, some students have difficulty figuring out the speed of the cart at point B because they are confused by the two different heights provided. These common difficulties are incorporated in the design of the multiple-choice questions. Table 2 presents the choices in algebraic form where all symbols have their standard meaning.

TABLE 2. The algebraic form for the choices in question 2 and the different scores assigned in the "weighted multiple-choice" simulation. The correct answer is indicated by the shaded background. Except for choice (d), the speed at point B ($v_B$) is calculated correctly using $\sqrt{2gh_1}$ in choices (b), (c) and (e).

| Choice | Question 2 | Score |
|---|---|---|
| (a) | $N = mg$ | 0.2 |
| (b) | $N = mg + m\frac{v_B^2}{r}$ | 0.8 |
| (c) | $N = m\frac{v_B^2}{r}$ | 0.7 |
| (d) | $N = mg - m\frac{v_B^2}{r}$, $v$ calculated using $\sqrt{2g(h_1 + h_2)}$ | 0.9 |
| (e) | $N = mg - m\frac{v_B^2}{r}$ | 1.0 |

The MC questions and the corresponding free-response questions were administered on an exam in two introductory physics course (with 185 and 153 students.) The Force Concept Inventory scores (pre-/post-instruction) indicate that students in these two courses are comparable (no statistically significant difference). The free-response questions were the same as their counterparts in the MC format except that there were no choices provided. Students' performance on the free-response questions was graded using rubrics, with a full score of 1 for each question. The rubrics incorporated students' common difficulties. Different partial scores were assigned based on the problem solving approach and the principles used. An example of a rubric can be found in [4].

To construct two types of "simulated" MC score from the answers students provided for the free-response questions, student responses were first binned into different categories by comparing and matching their answers to the different choices in the MC questions. For the dichotomous MC simulation, a score of 1 (correct choice) or 0 (incorrect choice) was then assigned for the various categories. For the weighted MC simulation, to simulate the partial credit which is usually awarded for a free-response question, we assigned partial credit to different binned responses based upon approaches students used for the free-response questions. The scores assigned to each of the choices in this "weighted" MC simulation are shown in Tables 1 and 2. The different weights for the choices reflect the different levels of understanding students display. The weights are commensurate with the rubrics used to grade the free-response questions.

To summarize, students who were given the free-response questions were graded using three different methods: using a rubric, using a dichotomous MC simulation, and using a weighted MC simulation. On the other hand, students who were given the MC questions were graded using two methods involving dichotomous or weighted scoring. Table 3 summarizes the different methods used to analyze student performance in the two courses. In order to compare student performance in the two courses, in each of the course, students were divided into groups 1 to 5 based on their overall performance on the final exam (with group 5 representing the group of students performing the best on the final exam, followed by those in group 4, etc.). For each group, students' average scores on each question were plotted to compare the trends in student performance in the two courses.

TABLE 3. Summary of Grading Methods in Classes.

|  | Course A (given free-response questions) | Course B (given multiple-choice questions) |
|---|---|---|
| Graded using a rubric | Yes | -- |
| Multiple-choice (dichotomous) | simulated (*) | Yes |
| Multiple-choice (weighted) | simulated (*) | Yes |
| (*): Student responses were first binned into different categories by comparing and matching their answers to the different choices in the multiple-choice format and then assigning a score as discussed in the text. | | |

## FINDINGS

TABLE 4. Percentage of students binned into different categories for simulated MC by comparing their free-response answers to the choices in both MC questions.

| Choice | Question 1 | % | Question 2 | % |
|---|---|---|---|---|
| (a) | $mgsin\theta$ | 28 | $N=mg$ | 9 |
| (b) | $\mu_k F_N$ | 7 | $N = mg + m\frac{v_B^2}{r}$ | 31 |
| (c) | $\mu_s F_N$ | 46 | $N = m\frac{v_B^2}{r}$ | 28 |
| (d) | $mgcos\theta$ | 3 | $N = mg - m\frac{v_B^2}{r}$ (v calculated using $\sqrt{2g(h_1+h_2)}$) | 6 |
| (e) | none of the above | 16 (*) | $N = mg - m\frac{v_B^2}{r}$ | 14 |
| -- | -- | -- | Other | 12 |
| (*): Both choice (b) and choice (c) was found in one students' free-response answer in this category. | | | | |

Table 4 presents the percentage of students who were binned into different categories by matching their free-response answers to the choices in the corresponding MC questions. We find that out of the 153 students involved, 84% and 88% of students' free-response answers could be matched to the a priori choices in the multiple-choice questions 1 and 2, respectively. Except for the mistake of using 1-D kinematics equations instead of the principle of conservation of mechanical energy to find the speed at point B in question 2 (which cannot be detected in the multiple-choice question because both methods yield the same numerical value for an option in the MC question) the findings suggest that a carefully designed research-based multiple choice question can reasonably reflect the distribution of common difficulties students have (as detected in their free-response answers).

Students' average performance on questions 1 and 2 is presented in Figure 3. The black and white labels are used to distinguish students in course B, who were given the multiple-choice questions and students in course A, who were given the free-response questions. In general, we find that the trends for students' performance in the two courses are similar regardless of the question format they were given. For example, comparing students' rubric-graded free-response performance in course A to students' dichotomous multiple-choice performance in course B, we find that in both courses, students who displayed a higher level of expertise on the final exam (e.g., students in groups 4 and 5) on average typically scored higher on questions 1 and 2 in both formats than those who did not perform as well on the final exam.

Comparing students' performance in course A to course B, we also find that there is a better correspondence between students' performance on free-response questions in one class and the multiple-choice questions in another class (shown in Figure 3) if partial credits are awarded for both types of questions. The reason free-response performance for students in one class has a better match with the weighted multiple-choice performance in the other class than the dichotomous multiple-choice performance is that the weights for the weighted MC performance were similar to those used in the rubric to score the free-response questions.

Table 5 presents the correlation coefficients between students' free-response performance (graded using the rubrics) and the simulated multiple-choice performance (both dichotomous and weighted) in course A. It shows that the correlation coefficient is always higher for weighted multiple-choice simulation. The correlation between free-response and simulated dichotomous MC performance is higher for the question with a single stronger distracter choice

(question 1) in MC compared to the question with several distracter choices (question 2) each of which represent different level of understanding.

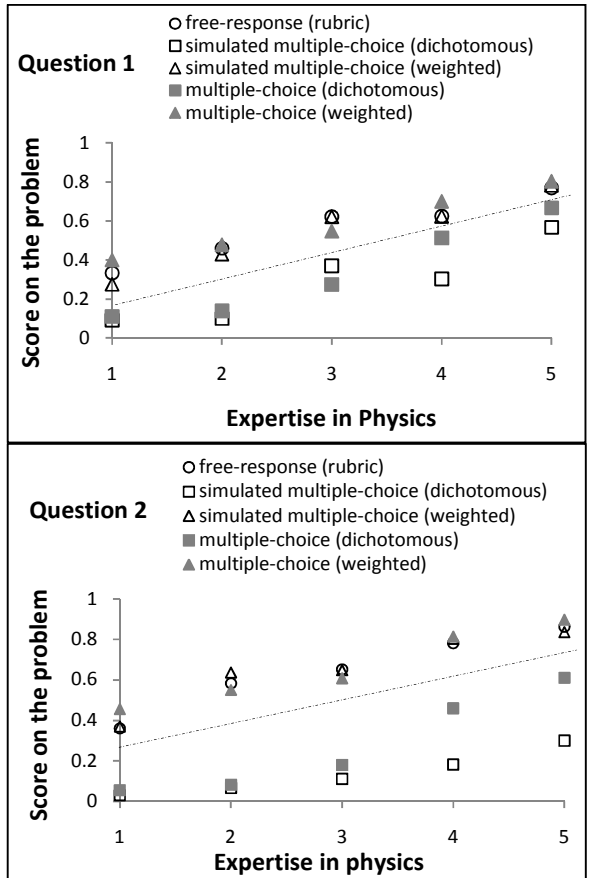

FIGURE 3. Students' average performance on questions 1 and 2. The white and black data labels are used to indicate students in course A (who were given the free-response question) and students in course B (who were given the multiple-choice questions), respectively. A dashed line is included on the figure to separate the data based for dichotomous case vs. the case where partial credits are assigned to the students.

TABLE 5. Correlation (N=153) between the free-response performance graded using the rubrics (FR) vs. the simulated multiple-choice performance for questions (Q) 1 and 2.

|  | FR vs. simulated multiple-choice (dichotomous) | | FR vs. simulated multiple-choice (weighted) | |
| --- | --- | --- | --- | --- |
|  | Q 1 | Q 2 | Q 1 | Q 2 |
| Correlation coefficient (r) | 0.890 | 0.483 | 0.928 | 0.945 |
| p-value | 0.000 | 0.000 | 0.000 | 0.000 |

## DISCUSSION

We find that the trends in student performance on the research-based multiple-choice questions given to one class (in which the distracter choices correspond to students' common difficulties) and the free-response questions given to another equivalent class are similar in that those who displayed a higher level of expertise on the final exam in each of the classes performed better on the questions than those who displayed a lower level of expertise regardless of the format of questions provided to them. Moreover, there is a good match between students' free-response answers in one class and the a priori choices in the MC questions administered to another class.

The findings suggest that research-based MC questions can reasonably reflect the relative performance of students on the free-response questions, especially if the answers for the MC questions are graded in a weighted manner by assigning partial credit to the different choices students selected. We note that (similar to the rubrics for the free-response questions) the weighting for the different alternative choices in the MC questions reflect the fact that some mistakes are not as "bad" as others despite the fact that they lead to students selecting an incorrect choice. In summary, if different scores are assigned to the different choices in the MC questions in the weighted model to reflect the different levels of understanding students display, there is a good overlap between students' MC performance in one class and the free-response performance in another class.

We re-emphasize that the fidelity of a MC question to a free-response performance depends strongly on the incorrect choices given [5]. If students' common difficulties found via research are incorporated, instructors can utilize MC questions without sacrificing accuracy in assessment of students' thinking processes. Free-response questions are useful only if they are graded carefully based on a good rubric. When they are graded leniently without a good rubric, the resulting scores will not typically reflect the appropriate level of understanding students' have. Weighted MC questions can be graded by a computer with weights corresponding to a good rubric for each distracter choice. Once the weights for the choices have been determined via research, MC questions can be as accurate for assessment purposes as rubric-based free-response questions without the time constraint.

## REFERENCES

1  C. Singh, AIP Conf. Proc. **951**, 196-199 (2007).
2  Salomon F. Itza-Ortiz, N. Sanjay Rebello, Dean A. Zollman, and Manuel Rodriguez-Achach, The Phys. Teach. **41**, 41-46 (2003).
3  C. Singh, Phys. Educ. **44** (5), 464-468 (2009).
4  S. Y. Lin and C. Singh, AIP Conf. Proc. **1289**, 209-212 (2010).
5  G. Aubrecht and J. Aubrecht, Am. J. Phys. **51** (7), 613-620 (1983).